\documentclass[prd,letterpaper,twocolumn,tightenlines,superscriptaddress,
psfig]{revtex4}

\usepackage{amsmath,amssymb,amsfonts,graphicx,dcolumn}
\usepackage{epstopdf}
\usepackage{color}
\usepackage[utf8]{inputenc}
\usepackage{amsthm}
\usepackage{fontenc}
\usepackage{tikz}
\usepackage[dvips]{hyperref}
\RequirePackage{slashed}
\usepackage{cancel}
\usepackage[normalem]{ulem}
\newcommand{\ie}{\textsl{i.e. }}
\newcommand{\eg}{\textsl{e.g. }}

\begin{document}


\twocolumngrid
\title{Relativistic effects in model calculations of double parton distribution 
functions}

\author{Matteo Rinaldi}
\affiliation{Instituto de Fisica Corpuscular (CSIC-Universitat de Valencia), 
Parc Cientific UV, C/ Catedratico Jose Beltran 2, E-46980 Paterna 
(Valencia), Spain.}

\author{Federico Alberto Ceccopieri}
\affiliation{ Dipartimento di Fisica e Geologia,
Universit\`a degli Studi di Perugia and Istituto Nazionale di Fisica Nucleare,
Sezione di Perugia, via A. Pascoli, I - 06123 Perugia, Italy}
\affiliation{IFPA, Universit\`e de Li\`ege, B4000, L\`iege, Belgium}	
\date{\today}

\begin{abstract}
\vspace{0.5cm}
In this paper we consider double parton distribution functions (dPDFs) 
which are the main non perturbative ingredients appearing
in the double parton scattering cross section formula in hadronic collisions.
By using recent calculation of dPDFs by means of constituent quark models 
within the so called Light-Front approach,
we investigate the role of relativistic effects  on dPDFs.
We find, in particular, that the so  called Melosh 
operators, which allow to properly convert 
the LF  spin into  the canonical one and incorporate a proper treatment 
of boosts, produce sizeable effects on dPDFs.
We discuss specific partonic correlations induced by these operators in 
transverse plane which are relevant to the proton structure and  
study under which conditions these results are stable against 
variations in the choice  of the proton wave function.
\end{abstract}

\maketitle
 
\section{\label{sec:intro}Introduction}
A proper description of final states in hadronic collisions 
requires the inclusion of multiple partonic interactions 
(MPI)~\cite{paver,Mekhfi,sjostrand}, 
\textsl{i.e.} a mechanism which takes into account the possibility
that more than one couple of partons may interact in a given hadronic 
collisions. 
This possibility emerges naturally since both colliding hadrons are extended
objects in transverse plane, at variance with processes involving point like
probes, as in Deep Inelastic Scattering
where, to date, no MPI effects have been reported. 
Multiple parton interactions enhance particle yields
at low transverse momenta, affecting multiplicities and energy flows.
MPI play an important role also in events characterized 
by an hard scale where they may contaminate the primary event with production of 
secondaries which contribute to the so called underlying event. 
In recent years, given the LHC operation, renewed interest 
has been paid to double parton scattering (DPS),
in which a couple of partons from each hadron interacts between each other.
If both interactions are hard enough, perturbative techniques can be 
applied and, as such, this class of processes need to be well controlled since
they might represent a background to New Physics Searches. At the same
time DPS has its own physical interest being sensitive to the nucleon structure. 
In particular, the cross section for this process 
depends on non-perturbative quantities, the so called double parton 
distribution functions (dPDFs). The latter encode the probability of 
finding two interacting partons, with longitudinal momentum fraction, w.r.t. the proton
one, $x$, and relative transverse distance $\vec b_\perp$, offering the opportunity to 
investigate parton momentum and spin correlations in the nucleon, unveiling new
information on the its structure, see Ref.~\cite{calucci}.
Since dPDFs are two-body distributions, this knowledge is complementary to the
one encoded in other type of 
(one-body) distributions, such as generalized parton distributions (GPDs) and
transverse momentum 
dependent distributions (TMDs).
To date,  dPDFs are very poorly known objects. Little guidance on their
structure come from sum rules which relate them to ordinary PDFs, see 
Refs.~\cite{gaunt,ceccopieri2}, 
while their perturbative QCD evolution is still debated due to the presence
of the so called inhomogeneous term in the evolution equations, see 
Refs.~\cite{snigirev03,diehl_1,diehl_2,ceccopieri1}.
In this situation it is clear that a proper theoretical modelization 
of DPS signal is quite challenging. This problem has been 
circumvented expressing the DPS cross section, $\sigma_{DPS}$, with final state 
$A+B$, by the following ratio, see \eg Ref. \cite{MPI15}:
\begin{eqnarray}
\sigma^{A+B}_{DPS}  = \dfrac{m}{2} \dfrac{\sigma_{SPS}^A 
\sigma_{SPS}^B}{\sigma_{eff}}\,,
\label{sigma_eff_exp}
\end{eqnarray}
where $m$ is combinatorial factor depending on the final states $A$ and $B$ 
($m=1$ for $A=B$ or $m=2$ for $A \neq B$) and
$\sigma^{A(B)}_{SPS}$ is the single parton scattering cross section with final 
state $A(B)$. The ratio in Eq.~(\ref{sigma_eff_exp}) 
relies essentially on the assumption  that the two hard scattering can be factorized,
an hypothesis which has been investigated in detail for the double 
Drell-Yan process in Ref.~\cite{DG_Glauber}. 
Furthermore dPDFs are often built up in a full factorized form of the type:
\begin{eqnarray}
\label{fact}
F_{a b}(x_1,x_2, b_\perp, Q^2) &\sim& f_a(x_1,Q^2) f_b(x_2,Q^2)
 \\  \nonumber  &\times& (1-x_1-x_2)^n T(b_\perp)~,
\end{eqnarray}
where $f_a(x_1,Q^2)$ and $f_b(x_2,Q^2)$ are the standard PDFs evaluated at
the scale $Q^2$, $n \ge 0$ is a phenomenological parameter which takes into 
account possible phase space effects on the kinematic boundary, see Ref.~\cite{gaunt},
and the function $T(b_\perp)$ captures parton correlations in the tranverse 
plane. This ansatz for dPDFs exploits the idea that, 
for decreasing parton fractional momenta, $x$, the parton population in the 
nucleon grows up, resulting in a substantial longitudinal decorrelation of the
joint distribution $F_{a b}$. The double parton interaction rate is then totally encapsulated 
in the function $T(b_\perp)$.
In such a factorized approach, the effective cross section 
appearing in Eq.~(\ref{sigma_eff_exp}) is simply given by
\begin{equation}
\sigma_{eff}^{-1} =  \int d^2b_\perp [T(b_\perp)]^2
\label{bprofile}
\end{equation}
and, by construction, does not show any dependence on parton fractional momenta, 
hard scales or parton species.
Due to the rather easy technical implementation of Eq.~(\ref{sigma_eff_exp})
and the almost total inclusiveness of the experimental analyses performed 
so far, all the present knowledge on DPS cross sections
has been condensed in the experimental and model dependent extraction of 
$\sigma_{eff}$~\cite{MPI15,S1,S2,S3,S4,S5,S6,S7}. 
To date, the corresponding number determined so far ($\sigma_{eff} \simeq$ 
15 mb) is compatible, within errors, 
with a constant, irrespective of centre-of-mass energy of the hadronic 
collisions and final state (A+B) considered.
Given this situation, many features of dPDFs are essentially unconstrained.
It is therefore clear that non perturbative methods may give access to some 
relevant properties on these 
distributions~\cite{manohar_2,noi1,noi2,ruiz,ruiz1,kas_dress}, allowing, 
for example, to establish to which extent such dPDFs models
may correctly reproduce the magnitudo of the transverse correlation encoded
in $\sigma_{eff}$, see for instance results of Refs.~\cite{nois, noiADS} on this 
point. 

In this work, starting from the results obtained in Ref. \cite{noi2},
where dPDFs have been calculated in valence region within a fully relativistic 
covariant treatment, the so called Light-Front (LF) approach,
we identify model independent effects induced on dPDFs by the relativistic 
treatment, in particular the violation of the factorized ansatz and the { effects} of 
parton correlation in the transverse plane { in the proton structure}. We also try 
to quantify the corresponding impact on observable-related quantities. 

The paper is organized as follows. In Section~\ref{sec:LF} we outline 
the structure of dPDFs and relativistic operators.
In Section~\ref{sec:HM} we describe the details of the hadronic
models used in the analysis.
In Section~\ref{sec:Results} we discuss the relevant issue 
of dPDFs factorization in longitudinal and transverse space
and the impact of the correct treatment of relativistic effects 
on dPDFs. We finally draw our conclusions in Section~\ref{Sec:Conc}.

\section{\label{sec:LF} The Light-Front approach and relativistic effects}

Following Ref.~\cite{noi2}, dPDFs have been calculated starting from their
Light-Cone correlator, which formally defines them in QCD. A suitable 
expression for dPDFs has been presented in Ref.~\cite{noi2}:
\begin{eqnarray}
 \label{dpdf}
\nonumber
F(x_1,x_2, \vec{k}_\perp) &\propto& \int d\vec k_1 d\vec k_2~\Psi 
\left(\vec k_1+ \dfrac{\vec k_\perp}{2}, \vec k_2- \dfrac{\vec k_\perp}{2}  
\right)
\\
&\times&\Psi^\dagger \left(\vec 
k_1- \dfrac{\vec k_\perp}{2}, \vec k_2+ \dfrac{\vec k_\perp}{2}  \right)
\hskip -1cm
\\
\nonumber
&\times& \delta  \left(x_1- \dfrac{k_1^+}{M_0} \right) \delta 
\left(x_1- \dfrac{k_1^+}{M_0} \right)
\\
\nonumber
&\times& \langle SU(6) | D_1^\dagger D_1  D_2^\dagger D_2 |SU(6) 
\rangle~,
\end{eqnarray}
where $\vec k_i$ is the intrinsic three-momentum of the $i-$ parton, 
$k_\perp$ is the relative transverse momentum between the two active partons,  
$\Psi$ is the proton wave function in momentum space and $|SU(6)\rangle $ is 
the spin-flavor state evaluated according the commonly adopted $SU(6)$ symmetry.
Here, as in Ref.~\cite{noi2}, a factorization between the spin and the 
spatial part of the proton wave function is assumed.
Let us remark that, thanks to this rigorous approach, the correct support of 
dPDFs is fulfilled, \ie the dPDF vanish in unphysical regions, \ie $x_1+x_2 >1$. 
A proper inclusion of relativistic effects is obtained 
via the so called Light-Front (LF) approach. This is a common procedure, 
largely used for the calculation non perturbative distributions, see \eg 
Refs.~\cite{Boffi1, boffi2, boffi3, boffitr,boffitmd, traini14}.
It should be noticed that dPDFs, calculated in momentum space, describe a 
system where two 
partons have a relative transverse momentum ($\pm \vec k_\perp$).
This unbalance physically arises since the difference of parton 
transverse momenta is not conserved between the amplitude and its complex conjugate~\cite{blok1,blok2}.
Due to this unbalance, the dPDFs are not densities in this representation and 
they can not be interpreted as probabilistic distributions.
In order to deal with distributions which admit a probabilistic 
interpretation we consider the Fourier transform of  Eq.~(\ref{dpdf}) 
w.r.t. to $\vec k_\perp$ which reads
\begin{equation}
\label{fourier}
F(x_1,x_2, \vec b_\perp) = \int \dfrac{d \vec k_\perp}{(2\pi)^2} e^{i \vec 
k_\perp \cdot \vec b_\perp} F(x_1,x_2, \vec k_\perp)~,
\end{equation}
with $\vec b_\perp$ being the relative transverse distance between the two 
partons.
In this paper we only consider the distribution of two unpolarized quarks of 
flavor $u$ (\textsl{c.f.r.} Refs. \cite{manohar_1, noi1,noi2}) 
so that $F(x_1,x_2, \vec k_\perp) \equiv F_{uu}(x_1,x_2, k_\perp)$ depends 
only on $k_\perp=|\vec k_\perp|$. Due to rotational invariance of $F_{uu}$ in case of 
unpolarized quarks, the Fourier transform reduces to:
\begin{equation}
\label{j0}
F(x_1,x_2,  b_\perp) = \dfrac{1}{2\pi} \int d k_\perp ~k_\perp  
J_0(b_\perp 
k_\perp) F(x_1,x_2,  k_\perp)~,
\end{equation}
being $J_0$ the Bessel function of the first kind and $b_\perp=|\vec b_\perp|$.
\begin{figure*}
\begin{center}
\includegraphics[scale=0.45]{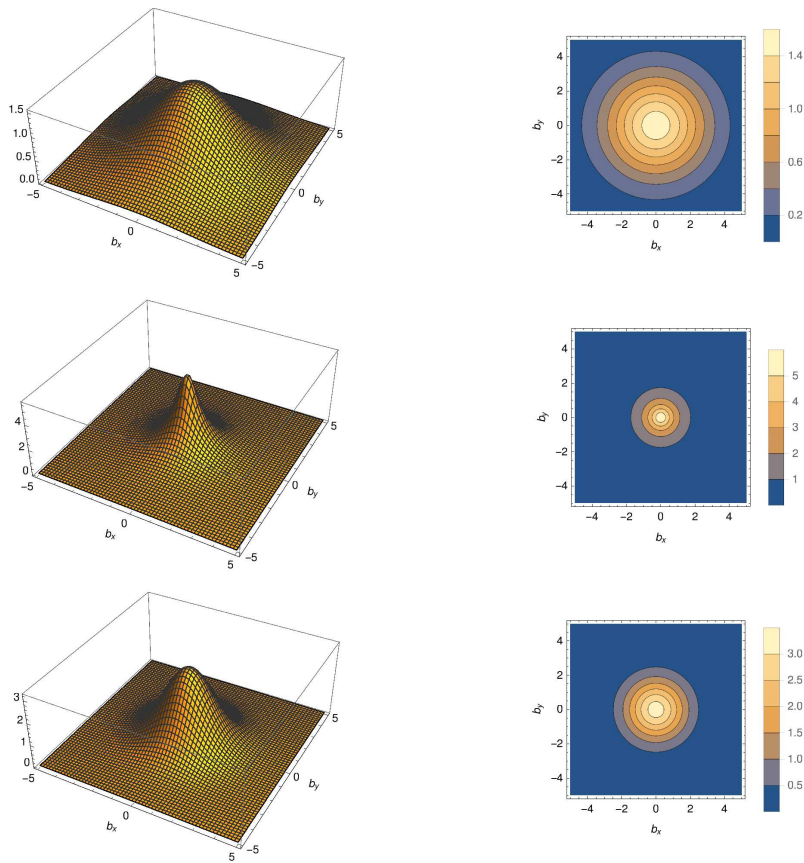}
\end{center}
\caption{\footnotesize  \textsl{Distribution evaluated via Eq.~(\ref{j0}}) by using 
different hadronic models presented in Section~\ref{sec:HM}, 
in particular the NR (top), the RL (middle) and HO$_{rel}$ (bottom) models at $x_1=0.2,~x_2=0.3$.}
\label{3d}
\end{figure*}
Anticipating some results discussed in the next Sections, we present in Fig.~(\ref{3d})  
the $b_\perp$-dependence of dPDFs at $x_1=0.2,~x_2=0.3$. 
The plots show the probability for two partons to initiate two separate
hard scattering as a function of their relative transverse distance, 
a unique information which is only accessible with such distributions.
The same distribution for dPDFs with longitudinal and transversally polarized 
quarks are likely to show departure from this symmetric structure 
giving access to new details of the proton structure. 
These spin effects on dPDFs are presently under investigation and will reported
in a separated paper. In Eq.~(\ref{dpdf}), the canonical proton wave function is 
calculated by means of constituent quark models (CQM). The 
LF proton wave function, which naturally arises in the 
LF approach, see \eg Ref.~\cite{Boffi1}, is related to the canonical one 
thanks to the introduction of the Melosh rotations~\cite{melosh} which appear 
in the last line of Eq.~(\ref{dpdf}).
The latter quantities are related to LF boosts which, in such an approach, are 
kinematical operators. Formally they are defined as 
\begin{eqnarray}
 \hat D_i = \dfrac{m+x_iM_0 +i(k_{i x} \sigma_y - k_{iy} \sigma_x    )  
}{\sqrt{(m+x_iM_0)^2+ k_{ix}^2+k_{iy}^2}}~,
\label{melo}
\end{eqnarray}
where $m$ is the constituent quark mass, $x_i$ the longitudinal momentum 
fraction carried by the $i$ quark, $\sigma_x$ and $\sigma_y$ are 
Pauli sigma matrices and $M_0$ is the energy that the proton had if quarks were
free and it depends itself on $\vec k_{i \perp}$ and $x_i$.
In particular, the Melosh operators are rotations  between the rest frame  
of the system reached through the Light-Front boost or canonical boost and allow 
to rotate Light-Front spin into the canonical one. For 
example a Light-Front state with momentum  $k$ and spin $\sigma$, $|k, \sigma 
\rangle_{LF}$ can be written  in term of canonical one, $|k, 
\mu \rangle_{IF} $  as follows :

\begin{eqnarray}
 |k, \sigma \rangle_{LF} \propto \sum_{\mu} \langle\mu|\hat D|\sigma \rangle~  
|k, 
\mu \rangle_{IF}~.
\end{eqnarray}
Thanks to this property, as pointed in Ref.~\cite{Boffi1}, one can 
convert the Light-Front proton wave function into the canonical one. This procedure 
is suitable for the calculations of non perturbative quantities, such as parton 
distributions, since  the proton wave function is usually evaluated by using the 
canonical Instant-Form approach.

From Eq.~(\ref{melo}), it is clear that
the structure of such operators induce non trivial correlations between the 
relevant variables 
at any energy scales. In order to visualize the effects produced by the term 
introduced in the last line of Eq.~(\ref{dpdf}), one can analytically evaluate:
\begin{eqnarray}
\label{Mel}
&&DD^\dagger (\vec k_\perp,x_1,x_2, \vec k_{1\perp}, \vec k_{2\perp}  ) = \\
 \nonumber
&&\langle SU(6) | D_1^\dagger D_1  D_2^\dagger D_2 |SU(6)\rangle ~.
\end{eqnarray}
However, since the dependence of the Melosh rotation w.r.t.  all the variables 
expressed in Eq.~(\ref{Mel}) is quite complicated, one can, without loss of generality,
examine its behavior in the limit $DD^\dagger 
(\vec k_\perp,x_1,x_2, \vec k_{1\perp}=0, \vec k_{2\perp}=0)$.
\begin{figure}
\includegraphics[scale=0.90]{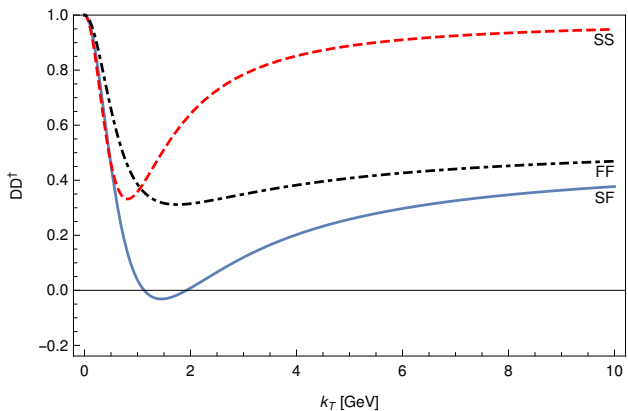}
\caption{\footnotesize{\textsl{The quantity  Eq.~(\ref{Mel}), as function of $k_T= 
k_\perp $,  evaluated in different regions of $x_1$ and $x_2$ 
with $\vec k_{1\perp}=\vec k_{2\perp}=0$.} }}
\label{Mel2}
\end{figure}
For this calculation, the allowed phase space $x_1+x_2 \le 1$, here and in the
following Sections,
is sampled in three different couples of points, which we found representative
for the effects we wish to discuss. In particular, we consider 
two fast partons (FF) with $x_1=0.2,~x_2=0.3$,
one slow and and one fast parton (SF) with $x_1=0.04,~x_2=0.3$ and
two slow partons (SS) with $x_1=0.04,~x_2=0.03$.
The calculation of Eq.~(\ref{Mel}) with these kinematic settings is
presented in Fig.~(\ref{Mel2})
where one may identify three distinct regions as a function of $k_\perp$.
For $k_\perp \rightarrow 0$ the Melosh's in all kinematic configurations 
reduce to unity. In an intermediate region 
of $k_\perp$ the curves show a dip whose depth depends on the chosen kinematic 
configuration and, in particular, becoming negative in the SF configuration. 
At larger $k_\perp$ the curves flattens out with different asymptotics. 
This complicated pattern, generated by Melosh's rotations, affects
the calculation of dPDFs, which, in general, are distributions
evaluated also at $k_\perp \neq 0$. 
It is worth noticing that such complicated behavior is due to the mixed 
combination of the four Melosh operators
combined with the proton spin structure assumed and described in its wave 
function.  These kind effects can not be observed in 
 known quantities such as standard PDFs or, e.g., in momentum distributions, 
being all these distributions depending on diagonal matrix elements (\ie evaluated at $k_\perp=0$).
In this case, in fact, the product of two Melosh reduces to the unity.
However, important effects due to the Melosh can be observed in model calculations of polarized PDFs. A crucial consequence of the presence of such operators  is the 
difference between the longitudinal and transversely polarized PDFs. In fact, 
since boosts commute with rotations in the non relativistic limit, such 
distributions are identical in this framework, see details on e.g.~\cite{boffitr}. Moreover, 
important effects are also found in the calculations of GPDs, see e.g. 
Refs.~\cite{Boffi1,boffi2, boffitr}, where some distributions are different 
from zero thanks to the presence of the Melosh. Such conclusions are also found 
in the analyses of TMDs in Light-Front CQM calculations Ref.~\cite{boffitmd}. 
Let us mention that Melosh effects can be appreciated in the calculation of nuclear 
spectral function and structure functions of the $^3$He within the Light-Front 
approach, see Refs.~\cite{spec,emc}. 

It should be noted, however, that the actual impact of Melosh on dPDFs 
is weighted by the chosen proton wave function and, in particular, 
by its structure at large parton momenta $k$. We address this issue 
in the next Section.

\section{\label{sec:HM} Hadronic Models}
The calculation of dPDFs via Eq.~(\ref{dpdf}) involves, 
beside the relativistic boosts just described, 
the modeling of the (canonical) proton wave function
which is obtained by means of CQM. 
The parameters of these models are fixed by comparison 
with subset of available data, \eg, the hadronic spectrum or 
the proton electromagnetic form factor at small momentum transfer.  
Since the aim of the present analysis is to identify potential model independent
effects on dPDFs, we consider a variety of proton wave functions.  The first 
model is the so called  Hypercentral 
quark model in both its relativistic (RL), Ref.~\cite{LF2}, and non 
relativistic (NR), Ref.~\cite{santop}, versions. 
\begin{figure}[t]
\includegraphics[scale=0.70]{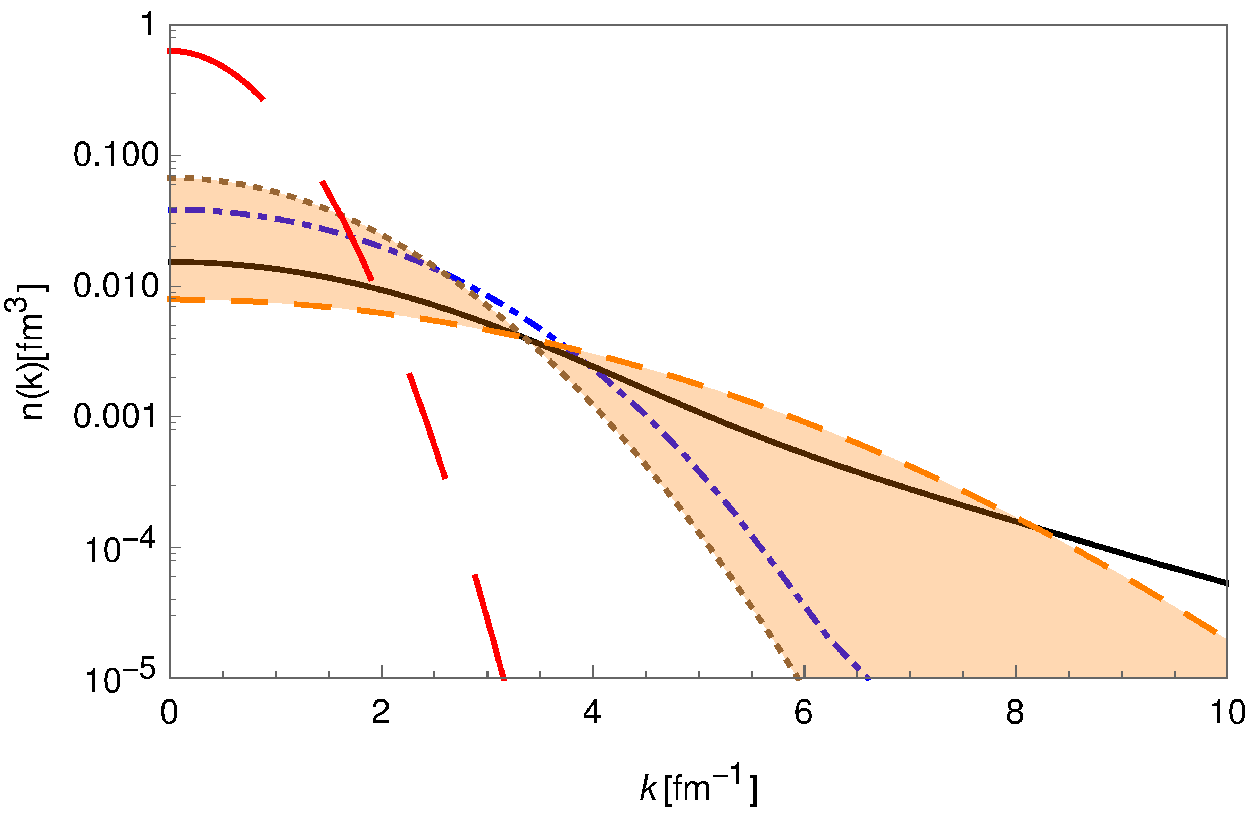}
\caption{\footnotesize{\textsl{Quark momentum distributions of the different models
used in this analysis. RL (full black), NR (dot-dashed blue), Original HO 
(dashed red), modified HO$_{nrel}$ (gold dotted),  modified HO$_{rel}$ (orange dashed). 
The orange band corresponds to variation of the parameter $6<\alpha^2<25 $ 
fm$^{-2}$.}}}
\label{1}
\end{figure}
The quark momentum distributions obtained within these two models are shown
in Fig.~(\ref{1}). The RL version (solid black) shows a broad tail extending 
at high momentum (hence relativistic) while the NR version (dot-dashed blue 
line) drops far more quickly at large momentum (hence non-relativistic). 
Since both versions assume a similar potential, 
we also consider a modified version of the harmonic oscillator (HO),
see details on the proton wave function calculated in such model in Ref. 
\cite{noi1}. In its original version, the width of the Gaussian 
structure of the proton wave function is fixed to $\alpha^2 = 1.35$ fm$^{-2}$.
As one can see in Figs.~(\ref{1}), the corresponding momentum distribution 
(red, long dashed line) decreases rather quickly w.r.t. $|\vec k|$, the quark 
momentum, showing a rather extreme non relativistic behavior, not suitable 
for the estimate of relativistic effects.
Given the relative mathematical simplicity of such a model,
we may construct a class of models of this type just 
varying the tunable parameter $\alpha$ in order  
to reproduce a  momentum distribution which can have either a 
relativistic or non relativistic behavior.
As shown in Fig.~(\ref{1}), we find that with the choice
$\alpha^2 = \alpha^2_{nrel}=6 $ fm$^{-2}$, we can simulate a non 
relativistic model (HO$_{nrel}$), while with the choice 
$\alpha^2 = 25$ fm$^{-2}$, the model (HO$_{rel}$) develops a quite broad 
relativistic tail.
Let us stress that for such values the agreement between HO model predictions
and available experimental data is lost.
\begin{figure}[t]
\includegraphics[scale=0.60]{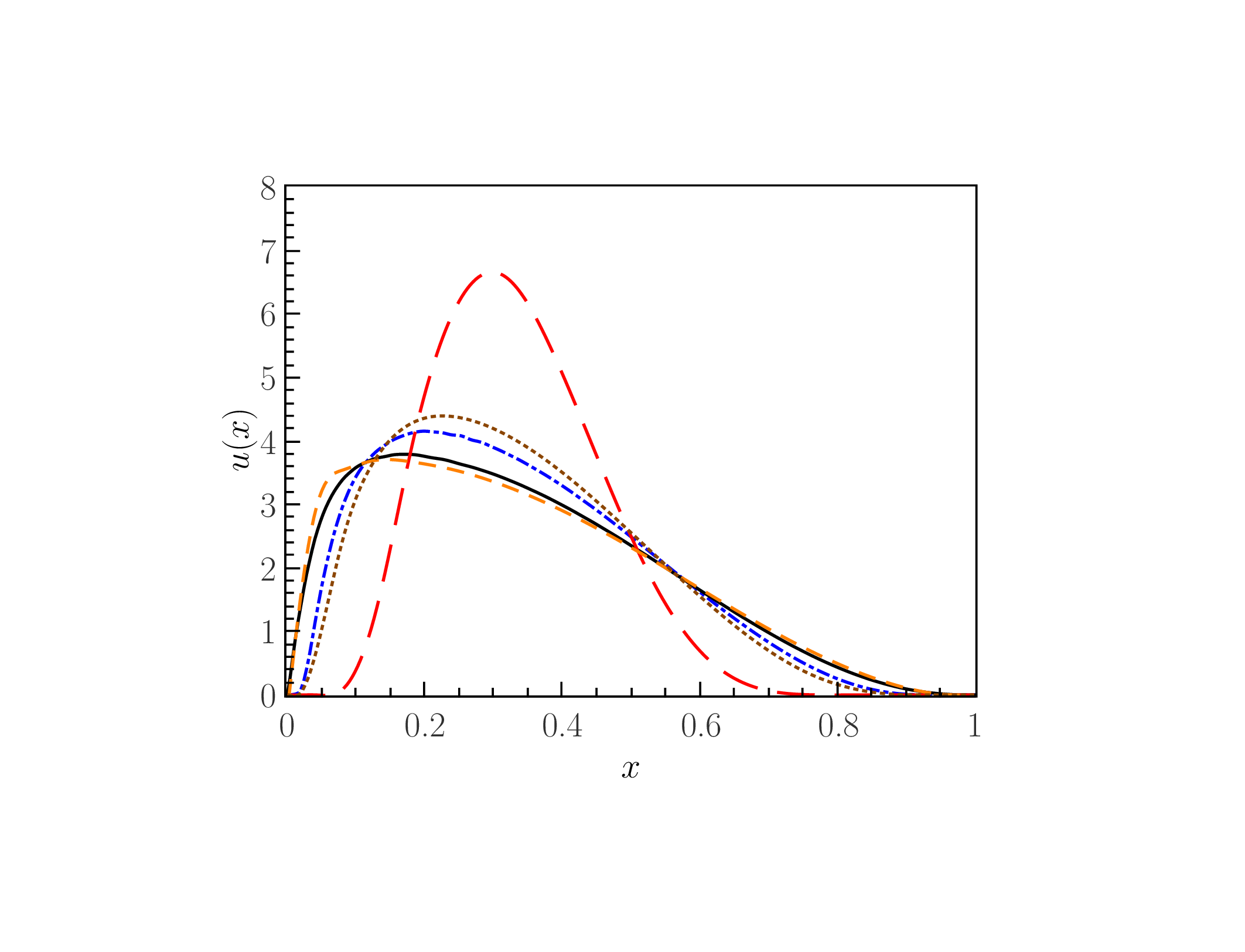}
\caption{\footnotesize{\textsl{Single parton distributions calculated with the 
different models used in this analysis.
RL full black line, NR dot-dashed blue line, Original HO 
dashed red line, modified HO$_{nrel}$ gold dotted line,  modified HO$_{rel}$ 
orange dashed line.}
}}
\label{2}
\end{figure}
We emphasize that the behavior of the CQM models, at large parton momentum,
determines the behavior at small $x$ of the corresponding 
parton distributions functions. 
This feature is easily explained considering the definition of the 
longitudinal momentum fraction carried by a quark in the LF approach:
\begin{equation}
 x_1 = \frac{k_1^+}{k_1^+ + k_2^+ + k_3^+}
 \label{x}
\end{equation}
where the light cone notation has been introduced, $a^+ = a^0 + a^3$ 
with $a^\mu$ being a generic four vector.
Since in Eq.~(\ref{x}) it is always $k^+ > 0$, the extreme small $x$ 
region can be achieved only if one parton has a very high momentum. 
Therefore a fast drop 
of quark momentum distributions at large $|\vec k|$, that is a non 
relativistic behaviour, determines a smooth vanishing of PDFs as $x\rightarrow0$.
On the contrary, PDFs corresponding to 
relativistic models still vanish in the limit $x\rightarrow0$ but with a much
harder behavior.
All these results are summarized in Fig.~(\ref{2}) where the $u$-quark 
distributions, obtained from all the considered models, are compared together.
With this selection of models, based on different potentials 
and showing different relativistic behavior, we now turn to the 
evaluation and discussion of dPDFs.

\section{\label{sec:Results}Calculations of  dPDFs}

\subsection{Breaking the factorized ansatz}
As already mentioned in the previous sections,
the present knowledge on the DPS cross section
is quite inclusive, being based 
on the approximations in Eqs. (\ref{sigma_eff_exp},\ref{bprofile}),
and it does not allow, at the moment, to pin down peculiar features of these 
distributions. 
\begin{figure}[t]
\begin{center}
\includegraphics[scale=0.65]{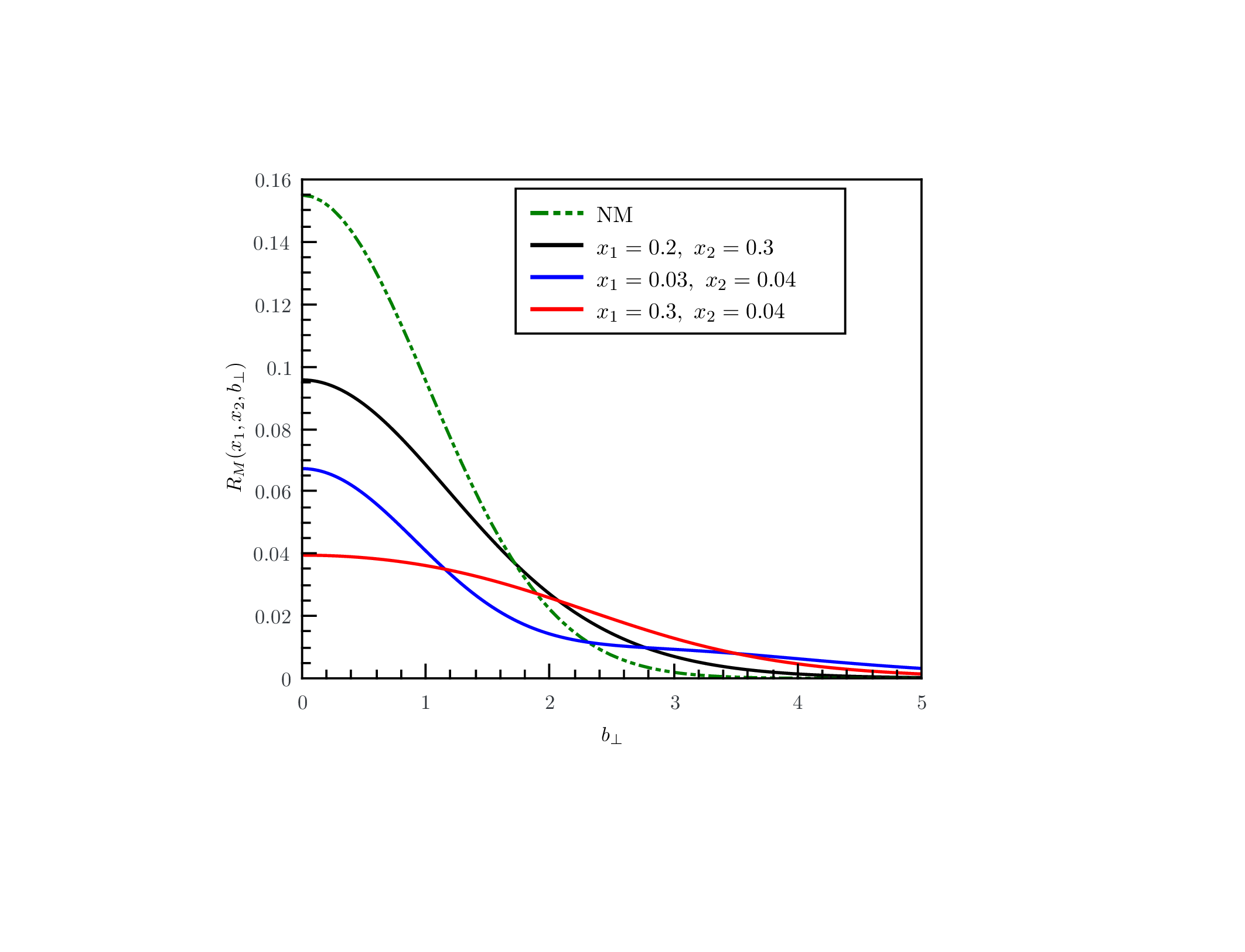} 
\end{center}
\caption{ \footnotesize \textsl{The ratio (\ref{RN}) evaluated using the 
HO$_{rel}$ model in three different regions of $x_1$ and $x_2$ as function of 
$b_\perp = |\vec b_\perp|$. In the legend the acronym ``NM'' specifies the 
calculation in which 
the Melosh rotations are neglected.}}
\label{RNN}
\end{figure}
Given this situation, the modeling of dPDFs aims 
to maximally exploit the current knowledge on the proton structure,
assuming a fully factorized form in all the relevant 
variables, as indicated in Eq.~(\ref{fact}).
\begin{figure*}
\begin{center}
\includegraphics[scale=0.44]{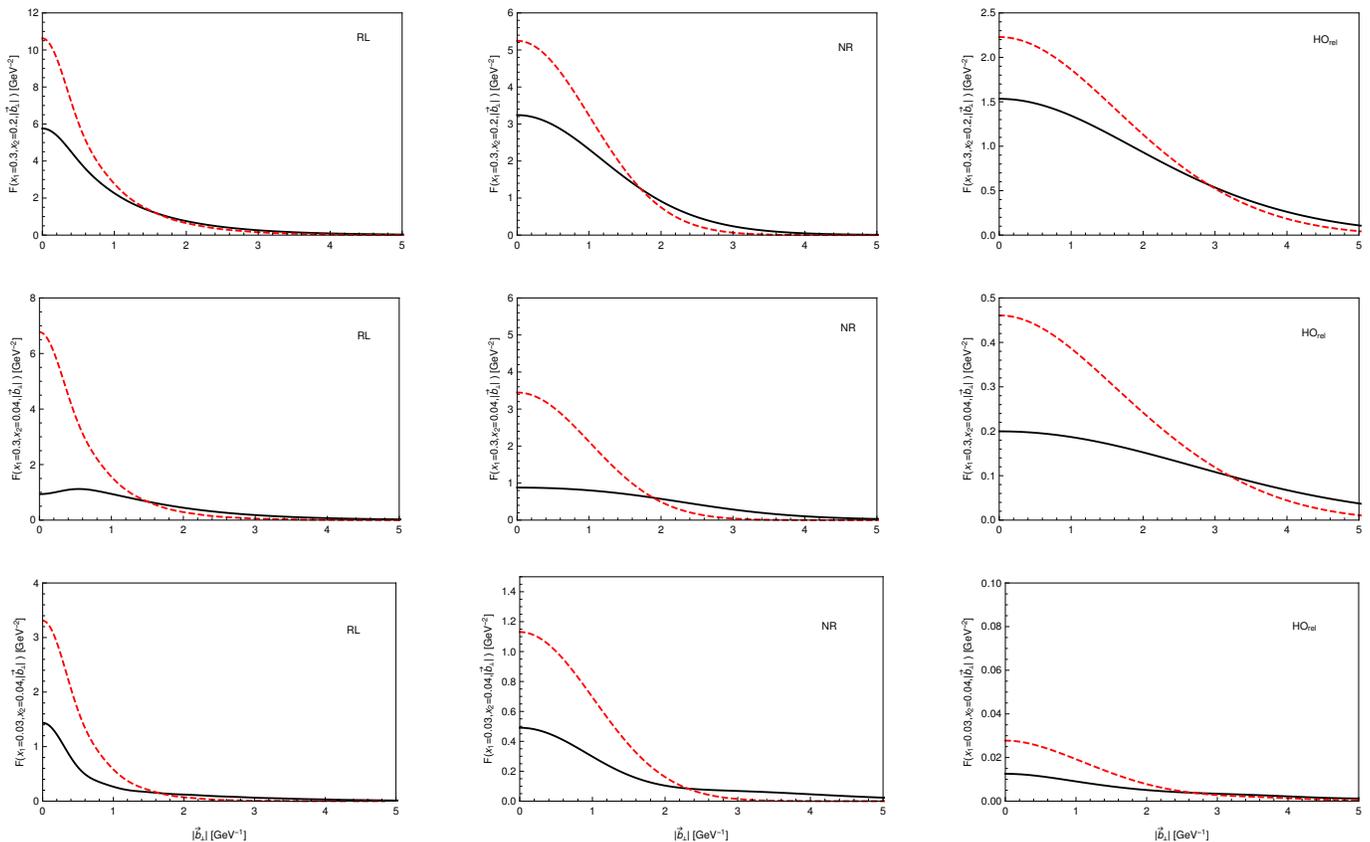}
\end{center}
\caption{\footnotesize  \textsl{The distribution Eq.~(\ref{j0}) 
as a function of $|\vec b_\perp|$ evaluated in three different 
kinematical configurations
(from top to bottom FF,SF,SS, respectively) using the relativistic model, the
NR one 
and the HO$_{rel}$ one (from left to right, respectively). Lines correspond 
to the evaluation of Eq.~(\ref{j0})
with (black) and without (red) Melosh's.}}
\label{dspacc_0203}
\end{figure*}
It is therefore clear that the modelization of dPDFs, beyond 
the approximation in Eq.~(\ref{fact}), will require more differential DPS 
measurements. 
Such a factorized ansatz is nevertheless valuable 
since, under a number of approximations, it allows to relate 
the double gluon distribution functions to the gluon GPDs,
exactly in the kinematic range of small $x$ where CQM model are difficult 
to extend and the corresponding theoretical predictions for $T(b_\perp)$ are 
lacking, see Refs.~\cite{blok1,blok2}. 
On the perturbative side, the evolution effects on such a factorized $b$-space ansatz  
have been investigated in Ref.~\cite{dkk}. 
In the moderate large values of fractional momenta, $x$, 
 natural domain of CQM with realistic potentials, as discussed in 
Refs.~\cite{manohar_1,noi1,noi2}, calculations show that both the 
factorization of dPDFs as a product of single parton distributions 
and, perhaps more interestingly, the  $(x_1,x_2)-k_\perp$ factorization
are violated. 
For recent results on the breaking of the factorization on the $x_1$ and 
$x_2$ dependence, see Ref.~\cite{noinew}.
Furthermore the breaking of the $(x_1,x_2)-k_\perp$ factorization might be  
generated both by the specific 
form of the proton wave function and by relativistic effects 
induced by Melosh operators. Within this context, 
the harmonic oscillator model appears to 
be particularly suitable to quantify to which extent such factorization 
breaking are due to relativistic effects alone.
Within this model, in fact, the $(x_1,x_2)-k_\perp$ dependences are entirely 
factorized, see Ref. \cite{noi1}.
In order to estimate these effects quantitatively, we evaluate the ratio
\begin{eqnarray}
\label{RN}
 R_N(x_1,x_2,b_\perp) = \dfrac{F(x_1,x_2, b_\perp)}{\int d\vec b_\perp~ 
F(x_1,x_2,b_\perp)}~.
\end{eqnarray}
It is worth to notice that, according to Eq.~(\ref{fourier}), $\int d\vec 
b_\perp~ 
F(x_1,x_2,b_\perp) = F(x_1,x_2,k_\perp=0)$ so that the denominator 
in Eq.~(\ref{RN}) does not depend on  the Melosh rotations which reduce to 
unity in the $k_\perp\rightarrow 0$ limit, see Fig.~(\ref{Mel2}).
The ratio, Eq. (\ref{RN}), calculated by using the HO$_{rel}$, is presented
in Fig.~(\ref{RNN}). The ratio $R_N$ with dPDFs evaluated 
without the Melosh rotations gives identical (superimposed) results in the three 
kinematic regions of $x_1$ and $x_2$, as expected. On the contrary, if Melosh rotation are 
taken into account, we observe a significant reduction 
of the magnitudo of dDPFs and a progressive broadening  of the $b_\perp$-dependence 
w.r.t. the distribution without Melosh depending on the partonal fractional momenta $x_i$. 
\begin{figure*}[t]
\begin{center}
\includegraphics[scale=0.43]{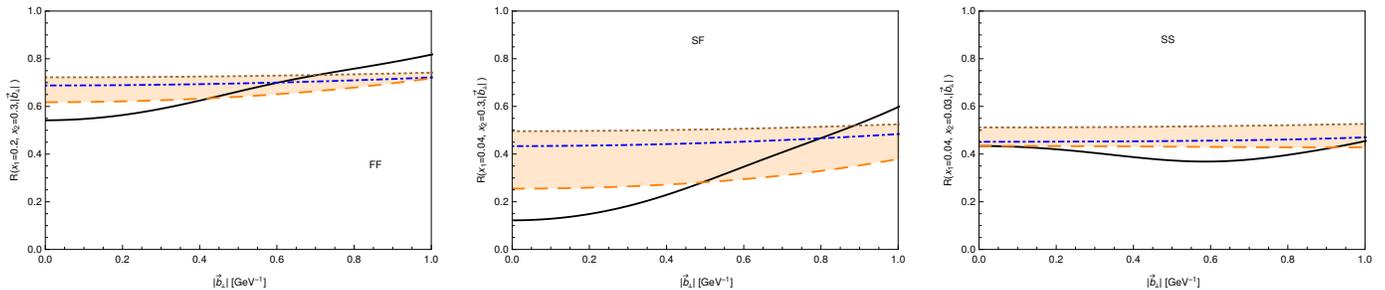}
\end{center}
\caption{\footnotesize{\textsl{The ratio Eq. (\ref{ratio}) evaluated as 
function of $|b_\perp|$ in different 
kinematic regions: $x_1=0.2,~x_2=0.3$ (left panel) and  $x_1=0.04,~x_2=0.3$ 
(middle panel) and $x_1=0.04,~x_2=0.03$ (right panel). RL full black line, 
NR dot-dashed blue line, Original HO dashed red line,
modified HO$_{nrel}$ gold dotted line,  modified HO$_{rel}$ orange dashed line. 
The orange band corresponds to variation of the parameter $6<\alpha^2<25 $ 
fm$^{-2}$.} }}
\label{ratiof}
\end{figure*}
This effect is sizeable especially in the SF configuration.
This observation leads us to conclude that for dPDFs evaluated through models
(which themselves may or may not show a $(x_1,x_2)-k_\perp$ factorization)
via Eq.~(\ref{dpdf}), at the hadronic scale, relativistic effects induce 
significant factorization breaking effects. In light of this result, it is 
worth to remark that possible modulation in the $b_\perp$-space, depending on 
parton fractional momenta, might not be disregarded altogether. 

\subsection{Relativistic effects}
It appears from the last Section that Melosh rotations
do not allow, in general, to factorize dPDFs in a longitudinal and transverse
distributions. More importantly they cause a significant 
reduction of the distributions, which, in turn, induce substantial 
variation of the corresponding DPS cross section.
In order to further investigate these effects, in this section 
we calculate dPDFs in $b_\perp$-space via Eq.~(\ref{j0}) with and without 
Melosh rotations. We note that in the latter case we basically reduce to the 
results presented in Refs.~\cite{manohar_2,noi1,noi2}.
The results of these calculations are shown in Fig.~(\ref{dspacc_0203}),
where predictions from different models are presented in columns
and different kinematical configurations in rows.
The $b_\perp$-spectra without Melosh rotation (NM) show a great variety in 
magnitudo and width, reflecting the difference in the used proton wave
functions.
In all cases the distributions are peaked at $b_\perp=0$ and show a finite 
behaviour in the short distance limit. 
If Melosh rotations are included (red dotted lines), we observe a
significant reduction of the magnitudo of the distributions. 
In the particular, in the SF kinematics, the magnitudo of the suppression 
is more pronunciated, and, for the RL models 
(left panel, middle row), the distribution tends to decrease
as $b_\perp\rightarrow 0 $ and it does show a maximum shifted to 
a non-vanishing value of $b_\perp$. We conclude that for relativistic models, 
in the SF region, these operators discourage the partons to be closed to 
each other. This model dependent behavior results from the combined effect of
the negative 
contributions of the Melosh rotations in Eq.~(\ref{j0}) and the large 
$|\vec k|$ tail of the RL model.
The amount of the suppression, induced by the Melosh operators, is again
conveniently quantified studying the ratio
\begin{eqnarray}
R(x_1,x_2, \vec b_\perp) = \dfrac{F(x_1,x_2, 
b_\perp)}{F_{NM}(x_1,x_2, b_\perp)}~, 
 \label{ratio}
\end{eqnarray}
where here $F_{NM}(x_1,x_2, b_\perp)$ is the distribution in Eq.~(\ref{dpdf}) 
once the Melosh rotations are neglected. The corresponding 
results are reported in Fig.~(\ref{ratiof}), where it is shown that 
the suppression slightly depends on the kinematical configurations, 
being smaller in the FF one and, on average, around 0.5 
in the FS and SS regions. Moreover such suppression is rather model 
independent, as can be inferred by the relatively contained spread of the 
orange band.
In all the previous sections, the main effects of the Melosh rotations have 
been analyzed directly on the dPDFs, either in momentum or coordinate space. 
It is worth to remark, however, that dPDFs appear in the DPS cross section 
in a convolution like formula which reads~\cite{paver}, 
\begin{equation}
\sigma_{DPS}^{A+B} \propto  \sum_{abcd} \int d\vec 
b_\perp~F_{ac}(b_\perp) \; F_{cd}(b_\perp) \;
\hat{\sigma}_{ab}^A \;  \hat{\sigma}_{cd}^B \;,
\label{DPS_cs}
\end{equation}
where we have neglected the dependences on longitudinal fractional 
momenta and $\hat{\sigma}$ are the elementary partonic cross sections
for the process $ab(cd)\rightarrow A(B)X$.
It is therefore clear that the details of the $b_\perp$ dependence
of the dPDFs gets obscured by the convolution and it is
intertwined with the dependences on longitudinal fractional momenta. 
Within this respect, we notice that a more direct access to the transverse
structure of dPDFs may be provided analyzing the DPS component in multijet photoproduction 
in $ep$ or $pp$ collisions. In this case the quasi-real photon, emitted either by the electron or the proton and fluctuating in $q\bar{q}$ dipoles, probes parton pair in the nucleon at a relative transverse distance of the order $b_\perp \sim 1/Q$~\cite{BS_photo,HERA_photo,Gaunt_splitting} 
while its low virtuality $Q$ can be controlled experimentally.
In the present context, a proper calculation of $\sigma_{eff}$ via eq.~(\ref{DPS_cs}) 
requires the selection of a definite final state $A+B$, the evaluation of corresponding partonic cross sections and the perturbative evolution  
of dPDFs from the hadronic scale (in the present work $\mu_0 = 0.1$  GeV$^2$, 
where only three valence quarks carry the proton momentum) to 
scales $\mu_A$ and $\mu_B$ characterising the hard processes.
We will report on these results in a separate publication.
In the present work, in order to get 
a quantitative estimate of the possible role of the Melosh
on observable-related quantities, following the lines of 
Refs.~\cite{nois, noiADS},
we define the ratio   
\begin{eqnarray}
R_{\sigma}(x_1,x_2) = \dfrac{\int  
d\vec 
b_\perp~F_{NM}(x_1,x_2, b_\perp)^2    }{\int  d\vec 
b_\perp~F(x_1,x_2, b_\perp)^2  }\,,
\label{ratios}
\end{eqnarray}
where the square is taken to mimic the analytic structure 
of the DPS cross section in  Eq.~(\ref{DPS_cs}) and Eq.~(\ref{bprofile}).
\begin{figure}[t]
\includegraphics[scale=0.70]{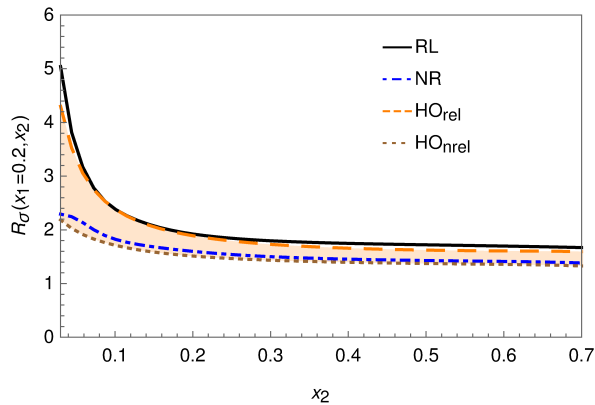}
\caption{\footnotesize{\textsl{The ratio $R_{\sigma}(x_1,x_2)$ calculated for 
fixed values $x_1$ as a function of $x_2$ within all the adopted models.}}}
\label{ratio2d}
\end{figure}
The ratio in Eq.~(\ref{ratios}) has been calculated by using the addressed
models in the three kinematics configurations. The results are presented in 
in Tab.~(\ref{Rsigmaratio}).
One should notice that in regions where the three CQM are 
completely different, the effects of the Melosh are rather independent on the 
choice of the detailed proton structure  considered. For 
the seek of transparency, some differences are found when small $x$ are
involved in the 
calculation. This feature can be seen as a limit of the present analysis. 
In fact, as already mentioned, the low $x$ region 
is associated to 
high momenta, where the three CQM substantially differ from each other and 
details of the models can not be totally separated by those of the 
relativistic treatment.
As shown in Fig.~(\ref{ratio2d}), the spread of $R_\sigma$ calculated 
within different models increases for decreasing $x_2$. This reflects 
different modelizations of the proton wave function at high quark momenta, see 
Fig.~(\ref{2}).
Nevertheless, it should be noticed that the value of the 
ratio and its spread, induced by different models, becomes constant
for, approximately, $x_2 > 0.1$. 
Therefore we may conclude that, in the valence region, the suppression factor
(a factor around 2) induced by Melosh rotations is quite a model independent
effect.
\begin{table}[t]
\begin{center}
\begin{tabular}{|l|l|l|l|l|} \hline
 & RL & NR & HO$_{rel}$ & HO$_{nrel}$ \\ \hline
 $R_{\sigma}(x_1=0.03,x_2=0.04)$ & 4.83 & 2.80 & 4.12 & 2.36  \\ \hline
 $R_{\sigma}(x_1=0.04,x_2=0.3)$   & 4.33 & 2.27 & 3.66 & 2.05  \\ \hline
 $R_{\sigma}(x_1=0.2,x_2=0.3)$    & 1.85 & 1.50 & 1.73 & 1.73  \\ \hline
\end{tabular}
\caption{\footnotesize{\textsl{The ratio $R_{\sigma}(x_1,x_2)$ calculated 
for different kinematical configurations and adopted models.}}}
\label{Rsigmaratio}
\end{center}
\end{table}

\section{\label{Sec:Conc}Conclusions}
In conclusion, in this work a quantitative analysis of relativistic effects on 
dPDFs has been provided thanks to the correct treatment of dPDF in a 
relativistic framework due to the LF approach, which implies the 
introduction of the so called Melosh rotations in order to achieve a full Poincar\`e covariant 
description of dPDFs.
We have discussed to which extent the Melosh rotations alone 
may affect the often assumed factorization in $(x_1,x_2)-k_\perp$ space, 
which is commonly used in the experimental analyses to extract the DPS cross 
section. We found that for relativistc models in the very low $x$ region, associated 
to high momenta of the quark, their effect is maximal and, for large unbalance of longitudinal 
momenta, they prevent the two interacting partons to be close to each other
in transverse space.
Then we have emphasized, by employing appropriate ratios, the role of such 
operators, and the degree of model independency of these results. 
In particular,  such rotations produce a strong reduction of the size of 
the dPDFs, w.r.t. the same calculation where such operators are neglected.  
As a consequnce they affect quantities related to $\sigma_{eff}$ by a suppression factor
which is, depending on the models, a factor ranging
from 2, in the FF configuration, to 4 in the SS one.
In closing, we have found that relativistic effects on dPDFs are sizeable 
and they should be always taken into account in these kind of model calculations.

\section*{Acknowledgements}
\noindent
This work is supported in part through the project 
``Hadron Physics at the LHC: looking for signatures of
multiple parton interactions and quark gluon plasma formation (Gossip project)",
funded by the ``Fondo ricerca di base di Ateneo" of the Perugia University.
This work was supported in part by the Mineco under contract 
FPA2013-47443-C2-1-P and SEV-2014-0398.

We warmly thank Sergio Scopetta, Vicente Vento, Marco Traini, Jonathan Gaunt and Alberto Accardi for 
many useful discussions.


\end{document}